%% file: conference_101719.tex
\documentclass[conference]{IEEEtran}
\IEEEoverridecommandlockouts
\usepackage[utf8]{inputenc}
\usepackage[T1]{fontenc}
\DeclareUnicodeCharacter{0307}{\textdotaccent}

\usepackage{amsmath,amssymb,amsfonts}
\usepackage{algorithm}
\usepackage{algpseudocode}
\usepackage{graphicx}
\usepackage{float}
\usepackage{subfigure}
\usepackage{textcomp}
\usepackage{xcolor}
\usepackage{booktabs}
\usepackage[numbers]{natbib}
\input{defs}

\def\BibTeX{{\rm B\kern-.05em{\sc i\kern-.025em b}\kern-.08em
    T\kern-.1667em\lower.7ex\hbox{E}\kern-.125emX}}
\begin{document}

\title{Data-Driven Risk Measurement by SV-GARCH-EVT Model}

\author{\IEEEauthorblockN{1\textsuperscript{st} Minheng Xiao*}
\IEEEauthorblockA{\textit{dept. Integrated System Engineering} \\
\textit{Ohio State University}\\
Columbus, USA \\
minhengxiao@gmail.com*}
% \and
% \IEEEauthorblockN{2\textsuperscript{nd} Minheng Xiao}
% \IEEEauthorblockA{\textit{dept. Integrated System Engineering} \\
% \textit{Ohio State University}\\
% Columbus, USA \\
% minhengxiao@gmail.com}
}
\maketitle

\begin{abstract}
This paper aims to more effectively manage and mitigate stock market risks by accurately characterizing financial market returns and volatility. We enhance the Stochastic Volatility (SV) model by incorporating fat-tailed distributions and leverage effects, estimating model parameters using Markov Chain Monte Carlo (MCMC) methods. By integrating extreme value theory (EVT) to fit the tail distribution of standard residuals, we develop the SV-EVT-VaR-based dynamic model. Our empirical analysis, using daily S\&P 500 index data and simulated returns, shows that SV-EVT-based models outperform others in backtesting. These models effectively capture the fat-tailed properties of financial returns and the leverage effect, proving superior for out-of-sample data analysis.
\end{abstract}


\section{Introduction}
Volatility is crucial in finance, especially for portfolio optimization and risk analysis. However, accurately estimating volatility is challenging because it varies over time (e.g., volatility clustering), often shows a negative correlation with asset returns (\textit{leverage effect}), and is typically not directly observable.

The negative correlation between stock returns and future volatility was first noted by \cite{black1976studies}. Since then, numerous volatility models have emerged, particularly in econometrics and financial mathematics. The ARCH model \citep{dhingra2024stock} and the GARCH model \citep{wu2024asymmetric}, which incorporates a moving average term, have been widely adopted. \citep{taylor2008modelling} developed the Stochastic Volatility (SV) model, which captures the dynamic nature of conditional variance more effectively. SV models include a white noise process to represent volatility changes \citep{kim1998stochastic, guo2024statistical}. Given that financial time series often deviate from normality and exhibit leverage effects, researchers have utilized non-normal conditional residual distributions such as the student’s t-distribution \citep{harvey1994multivariate}, Generalized hyperbolic skew Student's t-distribution \citep{ivanov2023stochastic}, and other skewed distributions \citep{harvey1994multivariate}.

Various estimation methods for stochastic volatility models have been proposed. Initially, the generalized method of moments (GMM) by \citep{hansen1982large} was used but found inadequate for small samples. Later efforts using the pseudo-maximum likelihood (QML) method with the Kalman filter also struggled with limited data. Bayesian methods, notably the Markov chain Monte Carlo (MCMC) technique~\citep{wang2023unbiased, gruffaz2024stochastic, bathina2023parameter}, have shown superior accuracy in parameter estimation. This paper employs the MCMC method, as \citep{jacquier2004bayesian} demonstrated its advantages over GMM and QML.

Since the Basel Accord II, Value at Risk (VaR) has become the standard for measuring market risk, estimating the maximum potential loss in a portfolio over a specified period at a given confidence level. VaR predictions depend on the distribution of financial returns and volatility forecasts. Yet, the dynamic nature of financial markets often undermines the typical assumption of a stable normal distribution \citep{ricca2024portfolio}. While various volatility models have been applied to estimate VaR, they frequently neglect fat tails, leverage effects, and extreme scenarios like the COVID-19 pandemic. For example, \cite{liu2018forecasting} applied HAR-related models with extreme value theory but overlooked leverage effects. \cite{afzal2021value} used DCC-GARCH to account for time-varying dynamics but ignored extreme cases. \cite{assaf2017stochastic} forecasted one-day-ahead VaR using SV and regime switch models without fully addressing the fat-tail characteristic of stock returns, leading to unexpected rejections in backtesting. Similarly, \cite{yang2017value} employed the MCMC method for estimating parameters in SV models with leverage but failed to consider extreme scenarios. Therefore, incorporating fat tails, leverage effects, and extreme cases is crucial for accurate VaR estimation.

This paper introduces a method that combines GARCH and SV models with MCMC for parameter estimation, integrating fat tails and leverage effects. We utilize daily S\&P 500 and simulated data to evaluate the algorithm’s convergence and performance in VaR estimation. Extreme value theory (EVT) is also applied to better address fat tails \citep{benito2023assessing}, and a goodness-of-fit test is conducted \citep{borner2023tail, an2024unleashing}. VaR forecasts and backtesting demonstrate that SV-EVT models are effective for VaR estimation. The SVtl-EVT model, which includes leverage effects and fat tails, outperforms others, although all models struggle with exceedance clustering. Further research is necessary to enhance model performance in extreme conditions, such as those experienced during the COVID-19 pandemic.

% This paper is structured as follows: Section 2 introduces three stochastic volatility models and their MCMC sampling procedures. Section 3 outlines extreme value theory and its application in risk measurement. Section 4 presents the estimation results of all models. Section 5 discusses backtesting methods and VaR backtesting results. Section 6 concludes the paper.

\section{The Stochastic Volatility Model}
\subsection{SV with linear regressors}
We begin by introducing the vanilla SV model with linear regressors. Subsequently, the analysis covers three generalized models: the SV model with Student's t errors (SVt), correlated errors (SVl), and their combination: Student’s t errors and leverage (SVtl). Let $\y_t = (y_1, \ldots, y_n)^\top$ be a vector of observations, the SV model is outlined as
\begin{gather}
y_{t} =\x_t^\top \boldsymbol{\beta}+\exp (h_{t} / 2) \varepsilon_{t}, \notag\\
h_{t+1} =\mu+\phi\left(h_{t}-\mu\right)+\sigma \eta_{t}, \notag \\
\varepsilon_{t} \sim \mathcal{N}(0,1) \text{ and } \eta_{t} \sim \mathcal{N}(0,1),
\label{eq:SV}
\end{gather}
where $\mathcal{N}(0,1)$ is the standard normal distribution and $\varepsilon_t$, $\eta_t$ are independent error terms. $\X=(\x_{1}^{\top}, \dots, \x_{n}^{\top})^{\top}$ is an $n \times K$ matrix including in its $t$-th row the vector of $K$ regressor at time $t$, $\h=(h_{1}, \ldots, h_{n})^{\top}$ represents the log-variance with $h_{0} \sim \mathcal{N}\left(\mu, \sigma^{2} /\left(1-\phi^{2}\right)\right)$ and  $\boldsymbol{\beta}=\left(\beta_{1}, \ldots, \beta_{K}\right)^{\top}$are regression coefficients. We denote $\boldsymbol{\theta} = (\mu,\phi,\sigma)$ as the SV parameters such that $\mu$ is the long-term level, $\phi$ is the persistence and $\sigma$ is the standard deviation of log-variance.

\subsubsection{SV with Student's t errors}
The basic model is restrictive for many financial series due to its tendency to exhibit fat tails. One extension of the basic model addresses this issue by allowing fat tails in the mean equation innovation. Formally, we have the error term
\begin{align}
\varepsilon_{t} & \sim t_\nu(0,1), 
\label{eq:2}
\end{align}
where $t_\nu (0,1)$ is the Student's t-distribution with $\nu$ degree of freedom, mean $0$ and variance $1$. The key difference between the SV and the SVt models is that in SVt, the observations follow a t-distribution. Additionally, as the degrees of freedom $\nu$ increase to infinity, the Student's t-distribution converges in law to the standard normal distribution.

\subsubsection{SV with leverage}
The basic model assumes a zero correlation between \(\varepsilon_t\) and \(\eta_t\), which can be extended to include the so-called \textit{leverage effect} by introducing correlation between the mean and volatility error terms. This enhancement involves adding a new parameter, \(\rho\), which indicates the correlation between an asset's returns and its volatility. A negative \(\rho\) suggests that a negative innovation in returns, \(\varepsilon_t\), is associated with increased contemporaneous and subsequent volatilities~\citep{jacquier2004bayesian}.
Eq.~\eqref{eq:SV} with a correlation between $\varepsilon_t$ and $\eta_t$ can be expressed as
\begin{align}
\label{eq:3}
\Sigma^\rho =
\begin{bmatrix}
1 & \rho \\
\rho & 1
\end{bmatrix},
\end{align}
hence, Eq.~\eqref{eq:SV} is a special case of Eq.~\eqref{eq:3} with $\rho = 0$.

\subsubsection{SV with Student's t errors and leverage}
\cite{jacquier2004bayesian} proposed to combine the t-error with the leverage effect, with Eq.~\eqref{eq:2} and Eq.~\eqref{eq:3} are generalized to
\begin{gather}
y_{t} =\boldsymbol{x}^\top_{t} \boldsymbol{\beta}+\exp \left(h_{t} / 2\right) \varepsilon_{t}, \notag\\
h_{t+1} =\mu+\phi\left(h_{t}-\mu\right)+\sigma \eta_{t}, \notag\\
\varepsilon_{t} \sim t_\nu(0,1) \text{ and } \eta_{t} \sim \mathcal{N}(0,1).
\label{eq:5}
\end{gather}
The correlation matrix of $(\varepsilon_t, \eta_t)$ corresponds to $\Sigma^\rho$ in Eq.~\eqref{eq:3}. The SVtl model is preferred as it captures the empirical observation that an increase in volatility typically follows a drop in stock returns. This model effectively accounts for increased fluctuations in returns through the negative correlation between the error terms for returns and volatility, known as the leverage effect.

\subsection{Estimation Methods}
\subsubsection{Markov Chain Monte Carlo}
MCMC combines Markov chain sampling with Monte Carlo estimation. It uses a Markov chain to draw samples \((\theta^{(1)}, \ldots, \theta^{(m)})\) from the posterior distribution \(p(\theta|x)\) and applies Monte Carlo methods to estimate the posterior mean \(E(g(\theta)|x)\) using the sample mean \(\overline{g(\theta)}\). Convergence of the chain occurs after several iterations with varying initial values. The initial \(k\) iterations, often non-smooth in distribution, are discarded in a process known as burn-in. The subsequent \(m - k\) iterations are then used for estimation.

Let $\boldsymbol{\theta} = (\phi, \sigma, \rho, \mu, \beta, \nu)$ be parameters of the SVtl, $\y = (y_1, \ldots, y_n)^\top$ be the stock returns, and $\h = (h_1, \dots, h_n)^\top$ be the unobservable log volatility. The conditional likelihood function of the model is
\begin{align*}
p(\y |\boldsymbol{\theta}, \h) = p(y_1, \dots, y_n | \boldsymbol{\theta}, h_1, \dots h_n) = \prod_{t=1}^n p(y_t | \boldsymbol{\theta}, h_t).
\end{align*}
The joint prior probability density of parameters $\boldsymbol{\theta}$ and the unobservable parameter $\h$ is then
\begin{align*}
p(\boldsymbol{\theta}, \h) &= p(\boldsymbol{\theta}, h_2, \ldots, h_n) = p(\boldsymbol{\theta}) p(h_0 | \boldsymbol{\theta}) \prod_{t=1}^n p(h_t | h_{t-1}, \boldsymbol{\theta}).
\end{align*}
The joint posterior probability density of $\boldsymbol{\theta}$ and $\h$ is proportional to the product of their prior probability and the conditional likelihood function as follows:
\begin{align*}
p(\boldsymbol{\theta}, \h | \y) \propto p(\boldsymbol{\theta}) p(h_0 | \boldsymbol{\theta}) \prod_{t=1}^n p(h_t | h_{t-1}, \boldsymbol{\theta}) \cdot \prod_{t=1}^n p(y_t | \boldsymbol{\theta}, h_t).
\end{align*}
In Bayesian methods, prior information improves parameter estimation accuracy. We follow \citep{jacquier2004bayesian} for the prior and posterior distributions as $\mu \sim \text{Normal}(0, 100)$, $(\phi+1) / 2  \sim \text{Beta}(5, 1.5)$, $\sigma^2 \sim \text{Gamma}(0.5, 0.5)$, $\nu \sim \text{Exponential}(0.1)$, $\rho \sim \text{Beta}(4, 4)$ and $\beta \sim \text{Normal}(0, 10000)$. For MCMC sampling, let $\boldsymbol{\theta} = (\phi, \sigma, \rho, \mu, \beta, \nu)$, $\y = \{y_t\}_{t=1}^n$, and $\h = \{h_t\}_{t=1}^n$. The prior distributions for \(\mu\) and \(\beta\) are given by
\begin{align*}
\mu \sim \mathcal{N}(\mu_0, \nu_0^2), \quad \beta \sim \mathcal{N}(\beta_0, \sigma_0^2).
\end{align*}
The MCMC sampling algorithm for drawing random samples from the posterior distribution of $(\boldsymbol{\theta}, \h)$ given $\y$ for the SVtl model follows \cite{jacquier2004bayesian}. After completing the sampling and achieving convergence, the SV model is built, and the estimated volatility and returns are transformed to standard residuals. This process accounts for fat tails and leverage effects, enabling the continuous fitting of extreme values. In the next section, we will introduce the extreme value theory and our dynamic risk measurement method.

\section{Extreme Value Theory and Dynamic Risk Measurements}
\subsection{Value-at-Risk}
Value at risk (VaR) represents the maximum potential loss of a portfolio of financial assets for a given confidence level $\alpha$. Let $P_t$ be the price of the financial assets at time $t$, and its log return at time $t$ is given by
\begin{align*}
Y_t = \ln\frac{P_t}{P_{t-1}} = \ln P_t - \ln P_{t-1}.
\end{align*}
Assume that the dynamic of \(Y\) is given by $Y_t = \mu_t + \sigma_t Z_t$, 
where the innovations $Z_t$ are a strict white noise process with mean 0 and variance 1. Let the density function of this return series be $f(y)$. The VaR at confidence level $\alpha$ can be expressed as
\begin{align*}
\text{VaR}_{\alpha}=-\inf \{y \mid f(Y \leq y)>\alpha\}.
\end{align*}
The formula for calculating the dynamic VaR of the return on assets $Y_t$, denoted as $\text{VaR}^t_\alpha$, is given by
\begin{align*}
\text{VaR}^t_\alpha = \mu_t + \sigma_t \cdot\text{VaR}_\alpha(Z_t),
\end{align*}
where $\mu_t$ is the return forecast at day $t$, $\sigma_t$ is the volatility forecast at day $t$, and $\text{VaR}_\alpha(Z_t)$ denotes the value-at-risk of the residual term $Z_t$ at $\alpha$-quantile.

Various methods have been proposed for estimating VaR, including parametric models that predict the return distribution of a portfolio. If this distribution is available in closed form, VaR simply corresponds to its quantile. For non-linear distributions, methods like Monte Carlo or historical simulation are used.

The parametric approach allows for updating factors through a volatility model. By selecting an appropriate distribution for the asset or portfolio, predicted volatility can define future return distributions. Consequently, the conditional predicted volatility measure \(\sigma_{t+\Delta}\) is used to estimate VaR for the next period, where \(\Delta\) represents the period length. This paper employs the historical simulation method for VaR estimation.

\subsection{Estimating Risk by Empirical Methods and GARCH model}
\subsubsection{Empirical Method}
Using the empirical distribution has been recognized as the simplest method to determine VaR. First, consider the empirical distribution $F^e_n$ with data points $\{l_i\}_{i=1}^n$, the empirical distribution places a mass of \(\frac{1}{n}\) at each \(l_i\), including repetitions. The VaR at confidence level $\alpha$, denoted as $\text{VaR}_\alpha(F^e_n)$, is given by:
\begin{align*}
\text{VaR}_\alpha(F^e_n) = l_{(\lceil n\alpha \rceil)},
\end{align*}
where \(\lceil n\alpha \rceil\) is the smallest integer \(k\) such that \(k \geq n\alpha\), where we sort the data points in ascending order as follows:
\begin{align*}
l_{(1)} \leq l_{(2)} \leq \ldots \leq l_{(n)}.
\end{align*}

\subsubsection{The GARCH Model}
The GARCH model combines a moving average component with an autoregressive component, which is expressed as
\begin{align}
\label{eq:GARCH}
&x_{t} = \sigma_{t} \epsilon_{t}, \notag\\
&\sigma_t^2 = \alpha_0 + \alpha_1 x^2_{t-1} + \beta_1 \sigma^2_{t-1}, \notag\\
&\epsilon_{t} \stackrel{i.i.d.}{\sim} \mathcal{N}(0, 1),
\end{align}
where $x_t$ is the log return series, and \(\alpha_0, \alpha_1 \geq 0\) to avoid negative variance. For inference, \(\epsilon_t\) is typically assumed to be normally distributed. However, given the fat-tailed property of financial return series consistent with the SV model, we use GARCH with Student's \(t\) innovation, \(\epsilon_t \sim t_\nu(0, 1)\). Table \ref{tab:garch} lists the results of parameter estimation by maximum likelihood estimation under our training data in Section \ref{Section:empirical}. According to the p-value, all parameters are significant.
\begin{table}[h]
 \caption{Parameter Estimation Results for GARCH}
  \centering
  \begin{tabular*}{\columnwidth}{ccccc}
    \toprule
    Parameters     &  Estimate&Standard Error  &p-value&Significance   \\
    \toprule
    $\alpha_0$ & $0.0433$ & $0.0109$ &  $7.01 \times 10^{-5}$&***\\
      $\alpha_1$ & $0.1749$ & $0.0300$ &  $5.54\times 10^{-5}$&***\\
        $\beta_1$ &$ 0.7847$ & $0.0309$ &  $2.00\times 10^{-5}$&***\\
    \bottomrule
  \end{tabular*}
  \label{tab:garch}
\end{table}

\subsection{EVT-POT}
It is common to assume asset returns follow a conditional normal distribution with time-varying variances for VaR calculation. However, this approach inadequately captures tail risks. To improve accuracy, VaR can be supplemented with Extreme Value Theory (EVT), which directly fits tail data to better estimate tail quantiles and address fat tails.

The Peaks Over Threshold (POT) in extreme value theory assumes the distribution function of the standard residual series \(\{Z_t\}\) is \(F(z)\). Given a threshold \(u\), the conditional distribution function \(F_u(y)\) of the random variable \(Z\) over the threshold \(u\) can be expressed as:
\begin{align}\label{eq:fu}
F_u(y) = F(y)(1 - F(u)) + F(u),
\end{align}
for \(0 \leq y < x_0 - u\), where \(x_0\) is the right endpoint of \(F\). According to \cite{pickands1975statistical}, for large classes of distributions $F$, there exists a positive function $\beta(u)$ such that: 
\begin{align*}
\lim_{u \to x_0} \sup_{0 \leq y < x_0 - u} \left| F_u(y) - G_{\xi, \beta(u)}(y) \right| = 0.
\end{align*}
When $u$ is sufficiently large, $F_u(y)$ can be approximated by the generalized Pareto distribution $G_{\xi, \beta}(y)$ as follows:
\begin{align}\label{eq:6}
G_{\xi, \beta}(y) =
\begin{cases}
1 - \left(1 + \xi \frac{y}{\beta}\right)^{-1/\xi} & \text{if }\xi \neq 0 \\
1 - e^{-y/\beta} & \text{if }\xi = 0,
\end{cases}
\end{align}
where $\xi$ is the shape parameter and $\beta$ is the scale parameter. Using the observations of \(\{Z_t\}\), \(\beta\) and \(\xi\) are estimated by maximum likelihood estimation. Let \(N_u\) be the number of samples exceeding threshold \(u\), then
\begin{align} \label{eq:7}
F(u) = \frac{N - N_u}{N}.
\end{align}
Substituting Eq.~\eqref{eq:6} and~\eqref{eq:7} into Eq.~\eqref{eq:fu}, the tail distribution \(\hat{F}(z)\) is:
\begin{align}\label{eq:Fz}
\hat{F}(z) =
\begin{cases}
1 - \frac{N_u}{N} \left[1 + \frac{\xi}{\beta}(z - u)\right]^{-1/\xi} & \xi \neq 0 \\
1 - \frac{N_u}{N} e^{-(z - u) / \beta} & \xi = 0.
\end{cases}
\end{align}
For a given confidence level \(\alpha\), by the definition of VaR and Eq.\eqref{eq:Fz}, we can obtain 
\begin{align*}
   \text{VaR}_\alpha(Z_t)= u + \frac{\beta}{\xi} \left(   \left(\frac{1-F(u)}{1-\alpha}\right)^\xi -1   \right), \quad \xi \neq 0
\end{align*}
Thus, the dynamic VaR model combining EVT, POT, SV, and GARCH is given by
\begin{table}[H]
\renewcommand{\arraystretch}{1.5}
\centering
\begin{tabular}{|c|l|}
\hline
\multicolumn{2}{|c|}{\textbf{SV(GARCH)-EVT-POT-VAR model}} \\
\hline
\textbf{SV Models}& 
\(\begin{aligned}
y_{t} &=\boldsymbol{x}^\top_{t} \boldsymbol{\beta}+\exp \left(h_{t} / 2\right) \varepsilon_{t}, \\
h_{t+1} &=\mu+\phi\left(h_{t}-\mu\right)+\sigma \eta_{t}, \\
\varepsilon_{t} & \sim t_\nu(0,1), \\
\eta_{t} & \sim \mathcal{N}(0,1), \\
\Sigma^{\rho} &=\left(\begin{array}{ll} 
1 & \rho \\ 
\rho & 1 
\end{array}\right)
\end{aligned}\) \\
\hline
\textbf{GARCH} & 
\(\begin{aligned}
x_{t} &=\sigma_{t,G} \epsilon_{t} \\
\sigma_{t,G}^2 &= \alpha_0 + \alpha_1 x^2_{t-1} + \beta_1 \sigma^2_{t-1,G} \\
\epsilon_{t} & \stackrel{i i d}{\sim}\left(0, 1\right)
\end{aligned}\)\\
\hline
\textbf{VaR} & 
\(VaR_\alpha^t = \mu_{t} + \sigma_{t}(u + \frac{\beta}{\xi} ( (\frac{1-F(u)}{1-\alpha})^\xi -1 ))\)\\
\hline
\end{tabular}
\end{table}
\noindent
where $\mu_{t}$ and $\sigma_{t}$ are the return and volatility forecasted at day $t$ from either SV models or GARCH model, and $\sigma_{t,G}$ is the standard deviation from GARCH model in Eq.~\eqref{eq:GARCH}. 

\subsubsection{The Threshold}
\begin{figure*}[h!]
    \centering
    \begin{tabular}{cc}
        \includegraphics[scale=0.42]{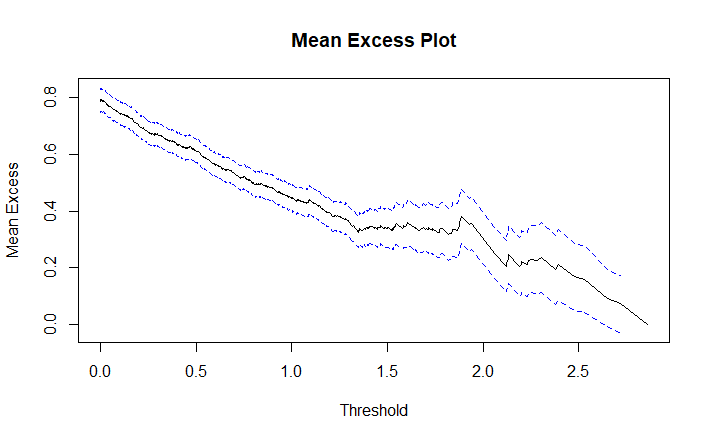} &
        \includegraphics[scale=0.42]{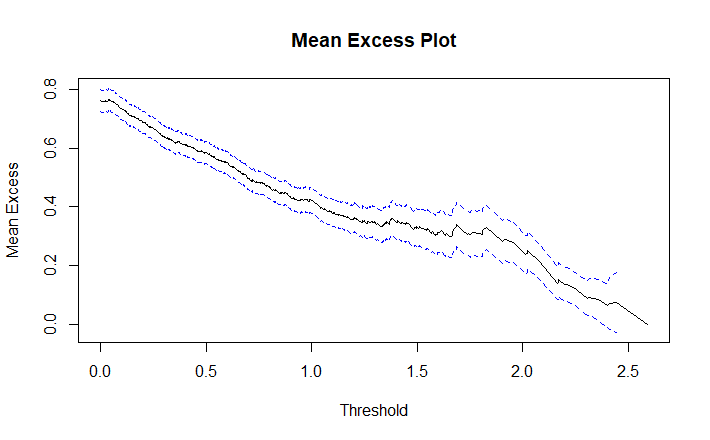} \\[-0.35cm]
        \includegraphics[scale=0.42]{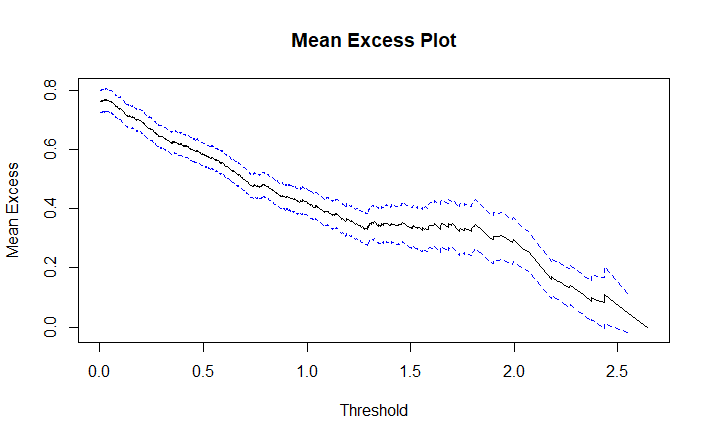} &
        \includegraphics[scale=0.42]{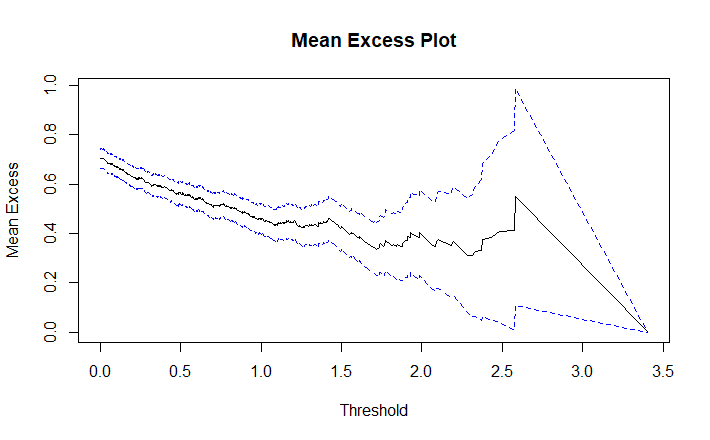}
    \end{tabular}
    \caption{Mean excess plots}
    \label{fig:exx}
\end{figure*}

To estimate the parameters of the POT model, we first select a reasonable threshold \(u\) and then estimate the parameters \(\xi\) and \(\beta\) using maximum likelihood estimation. A high threshold results in too little excess data, increasing the variance of the estimates, while a low threshold increases accuracy but introduces bias. Selecting the optimal threshold remains an unsolved problem in extreme value theory. The threshold is estimated using the mean excess function method denoted as
\begin{align*}
e(u) = E(X - u \mid X > u) = \frac{1}{n} \sum_{i=1}^n (x_{(i)} - u),
\end{align*}
where \(x_{(1)} < x_{(2)} < \ldots < x_{(n)}\) is the curve formed by the excess mean graph for point \((u, e(u))\). By selecting an appropriate threshold \(u\), \(e(x)\) is approximately linear for \(x \geq u_0\). Figure \ref{fig:exx} and Table \ref{tab:ex} present the estimates for the threshold and parameters using the excess mean method and maximum likelihood estimation (MLE). To assess the adequacy of the Peaks Over Threshold (POT) method, we apply the goodness-of-fit test, which calculates the Cramer-von Mises ($W^2$) and Anderson-Darling ($A^2$) statistics and their respective p-values. The model is considered appropriate if both P-values exceed 0.1, as detailed in Table \ref{tab:gof}.
% If the graph is upward-sloping for \(x \geq u_0\), the data shows a GP distribution with positive shape parameter \(\xi\), indicating a fat-tailed distribution. A downward-sloping graph indicates a thin-tailed distribution with \(\xi < 0\), and a horizontal line indicates an exponential distribution with \(\xi = 0\).

\begin{table}[h]
 \caption{Threshold for different models}
  \centering
  \begin{tabular*}{\columnwidth}{@{\extracolsep{\fill}}ccccc}
    \toprule
   & SVt     & SVl  & SVtl  & GARCH   \\
    \midrule
   $u$ & 2.122 & 2.394 & 2.403 & 2.365 \\
   $\xi$ & 0.385 & 0.354 & 0.435 & 0.744 \\
   $\beta$ & 0.085 & 0.048 & 0.061 & 0.148 \\
    \bottomrule
  \end{tabular*}
  \label{tab:ex}
\end{table}

\begin{table}[h]
 \caption{Goodness-of-fit testing for different models}
  \centering
  \begin{tabular*}{\columnwidth}{@{\extracolsep{\fill}}ccccc}
    \toprule
   & SVt     & SVl  & SVtl  & GARCH   \\
    \midrule
   $W^2$ & $0.0602^*$ & $0.0292^*$ & $0.0342^*$ & $0.0399^*$ \\
   $A^2$ & $0.4077^*$ & $0.3200^*$ & $0.3366^*$ & $0.3504^*$ \\
    \bottomrule
  \end{tabular*}
  \label{tab:gof}
\end{table}
All p-values are greater than 0.1, indicating that the POT model fits the tail data well. This confirms that the threshold selection is appropriate, making the use of the POT model for fitting tail data and VaR analysis of the SV and GARCH models reasonable.

\section{Empirical Results \label{Section:empirical}}
\subsection{Data Analysis}
We analyze the behavior of the S\&P 500 using training data from January 4, 2011, to December 30, 2016, and test data from January 3, 2017, to December 31, 2020, sourced from the WRDS dataset. According to \citep{french1987expected}, dividends were not adjusted in the stock index prices and reportedly had minimal impact on the estimation results. Returns are calculated using $y_t = 100 \times \ln (p_t/p_{t-1})$, where $p_t$ denotes the index value on day $t$. Summary statistics are provided in Table \ref{tab:1}.
\begin{table}[h]
 \caption{Summary statistics of daily returns of S\&P500}
  \centering
  \begin{tabular*}{\columnwidth}{@{\extracolsep{\fill}}ccccc}
    \toprule
    Mean & S.D & Skewness & Kurtosis & J.B. \\
    \midrule
    $0.037$ & $0.949$ & $-0.510$ & $4.504$ & $1346.570$ \\
    \midrule
    Ljung-Q($5$) & ADF & $\text{ACF}_1$ & $\text{ACF}_2$ & $\text{ACF}_3$ \\
    \midrule
    $28.309$ & $-12.017$ & $-0.044$ & $0.028$ & $-0.080$ \\
    \bottomrule
  \end{tabular*}
  \label{tab:1}
\end{table}

\begin{figure}[H]
\centering
\includegraphics[scale=0.4]{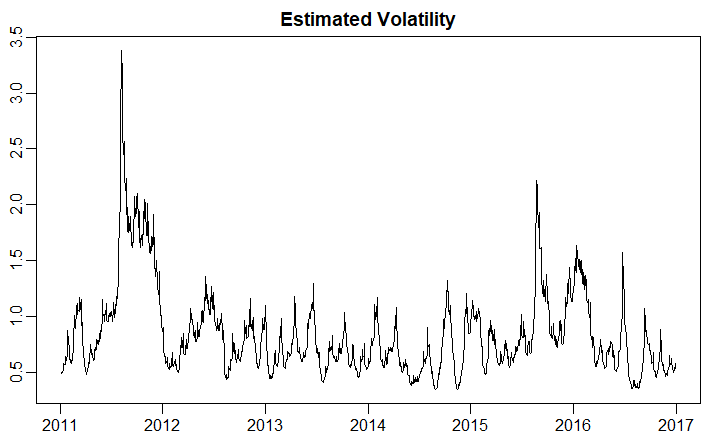}
\caption{Daily log-returns of S\&P 500 and Estimated Volatility}
\label{fig:estvoll}
\end{figure}

The table shows that the return series is negatively skewed, with frequent minor gains and occasional significant losses. It is leptokurtic, exhibiting a kurtosis of 4.504, which suggests a peaked distribution with fat tails, confirmed by Figure \ref{fig:estvoll}. From 2011 to 2012, the maximum daily loss was approximately 6.734. The ADF statistic of -12.017 confirms the returns are stationary with no unit root, and the J.B. statistic of 1346.57 indicates a deviation from normal distribution, aligning with observed skewness and kurtosis. Lastly, the Ljung-Box statistic suggests minimal serial correlation.

\subsection{MCMC Estimation Results}
\begin{figure*}[h]
\centering
\includegraphics[scale=0.9]{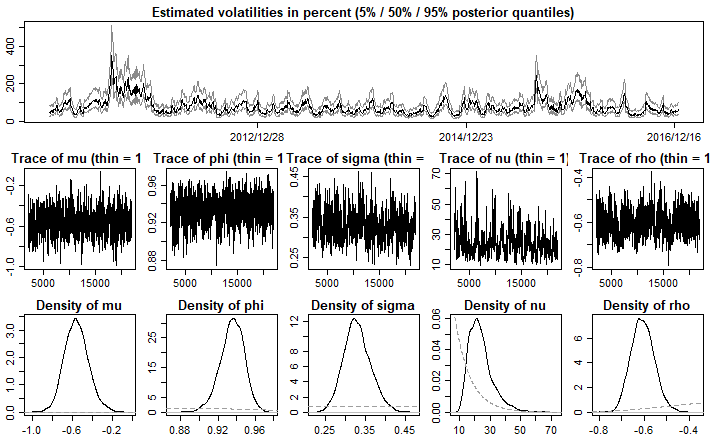}
\caption{Estimation results of the SVtl model for S\&P 500 data}
\label{fig:tracesvtl}
\end{figure*}

Bayesian analysis primarily focuses on assessing the joint posterior distribution of model parameters and latent quantities through summary statistics and visualizations of marginal posterior distributions. Table \ref{tab:mcmc} details the posterior means and standard errors for three stochastic volatility models. The negative leverage effect $\rho$ suggests asymmetry in the leverage effect~\citep{yu2005leverage}. All three models exhibit strong persistence in, approaching 0.94. Notably, $\mu$ increases from the SVt to SVtl model, indicating an upward adjustment in the long-run log-variance level when accounting for leverage effects and fat tails.
\begin{table*}[h]
 \caption{MCMC Estimation Result: Posterior Mean and Standard Error}
  \centering
  \begin{tabular*}{2\columnwidth}{@{\extracolsep{\fill}}cccccccccc}
    \toprule
 & \multicolumn{3}{c}{SVt} & \multicolumn{3}{c}{SVl} & \multicolumn{3}{c}{SVtl}\\ 
  \cmidrule(r){2-4}    \cmidrule(r){5-7}   \cmidrule(r){8-10}   
Parameters & Mean     & SD     & 95\% CI & Mean &  SD& 95\% CI & Mean& SD& 95\% CI\\
    \toprule
   $ \mu $&  -0.58 &0.15      &(-0.81,-0.31) &-0.57 &0.11&(-0.77,-0.40)&-0.56&0.12&(-0.75,-0.36)\\
    $\phi $   &0.94 &0.01     &(0.92,0.97)  &0.93 &0.01&(0.91,0.95)&0.94&0.01&(0.91,0.95) \\
 $  \sigma$     &0.30     &0.04   &(0.25,0.36) &0.34 &0.03 &(0.29,0.40)&0.33&0.03&(0.28,0.39)\\
    $\nu$    &21.96       &11.50   &(10.88,42.92) &  & & &24.00&8.00&(13.96,39.41)  \\
    $  \rho$     &        &   & &-0.70 &0.05 &(-0.77,-0.60)&-0.61&0.05&(-0.69,-0.52)\\
     $ \exp(\mu/2)$     & 0.76 &0.06   &(0.67,0.85) &0.75 &0.04 &(0.68,0.82)&0.76&0.05&(0.69,0.84)\\
    $  \sigma^2   $  & 0.09       &0.02   &(0.06,0.13) &0.12 &0.02 &(0.09,0.16)&0.11&0.02&(0.08,0.15)\\
      $\beta $    & 0.08       &0.02   &(0.05,0.11) &0.04 &0.02 &(0.01,0.06)&0.05&0.02&(0.02,0.07)\\
    \bottomrule
  \end{tabular*}
  \label{tab:mcmc}
\end{table*}
Figure \ref{fig:tracesvtl} displays the posterior daily volatility (in percent) $100 \times \exp(\boldsymbol{h}/2)$ with its median (black) and 5\% and 95\% quantiles (gray). The other panels summarize the Markov chains for the parameters $\mu, \phi, \sigma, \nu,$ and $\rho$. Specifically, trace plots are shown in the middle row, and the bottom row compares prior (gray, dashed) and posterior (black, solid) densities. The sampling process involved 20,000 MCMC draws after a 2,000 iteration burn-in, achieving convergence for all parameters.

\section{Backtesting}
In this section, we introduce several metric for the backtesting under our dynamic VaR estimation approach. 
We first provide the detail of what is called \emph{binomial approach}. Given the time series of past ex ante VaR forecasts and past ex post returns, the "hit sequence" of VaR violations can be defined as:
\begin{align*}
I_{t+\Delta}=
\begin{cases}
1, & \text{if } R_{t+\Delta} < -\text{VaR}_{t+\Delta}^{p} \\
0, & \text{if } R_{t+\Delta} > \text{VaR}_{t+\Delta}^{p}.
\end{cases}
\end{align*}
The ``hit sequence'' \(I_{t+\Delta}\) indicates VaR violations, returning 1 when the loss exceeds the forecasted VaR on day \(t+\Delta\), and 0 otherwise. For backtesting, we compute the sequence \(\{I_{t_0+j\Delta}\}_{j=1}^J\) over \(J\) days to count the number of violations denoted as
\begin{align*}
N^J_{t_0} = \sum^J_{j=1} I_{t_0 + j\Delta},
\end{align*}
which follows a binomial distribution. Define the statistic $\hat{Z}$ as follows:
\begin{align*}
\hat{Z} := \frac{\hat{N}_{t_0}^J - J(1-\alpha)}{\sqrt{J\alpha (1-\alpha)}},
\end{align*}
with the central limit theorem (CLT), the statistic $\hat{Z} \sim \calN(0,1)$ with  $1 - \beta$ confidence interval (CI) given by
\begin{align*}
\mathbb{P}[\tau_{-}^\beta \leq \hat{N}_{t_0}^J \leq \tau_{+}^\beta] \approx 1 - \beta,
\end{align*}
where $\tau_{\pm}^\beta := J(1-\alpha) \pm z_{1-\beta / 2} \sqrt{J \alpha (1-\alpha)}$ and $z_\gamma = N^{-1}(\gamma)$.

Additionally, the backtesting of our VaR model incorporates three test statistics from \cite{christoffersen1998evaluating}: the \emph{Unconditional Coverage Test} ($LR_{uc}$), which checks if the model predicts the correct frequency of exceedances; the \emph{Independence Test} ($LR_{ind}$), which verifies that exceedances are not clustered over time, and the \emph{Conditional Coverage Test} ($LR_{cc}$), which assesses both the frequency and independence of exceedances.

\subsection{Results}
Table \ref{tab:back} presents backtesting results. An asterisk (*) indicates model exceedances within the confidence interval, while double asterisks (**) denote rejection of the null hypothesis. The significance level for one-day VaR is set at 5\%, using 252 rolling windows. For $LR_{uc}$, a significant result confirms that expected and actual observations below the VaR estimate are statistically equivalent. However, rejection of the null hypothesis across most models, including empirical methods, suggests inadequate VaR accuracy. The dynamic VaR model proposed in this paper is validated by all but the empirical method based on $LR_{uc}$ results at 5\% significance. Conversely, all models fail the $LR_{ind}$ and $LR_{cc}$ tests, except the SVtl-EVT model, which shows minimal exceedance clustering with the lowest test statistics of 6.7 (critical value: 5.991). The period under review includes the late 2020s, marked by the COVID-19 pandemic's impact on stock markets. Notably, except for the empirical method, all model exceedances lie within the 95\% confidence interval, with the SVtl model demonstrating a suitable exceedance count of 38. This suggests that SV models integrated with EVT are viable, with SVtl performing best during this period. According to simulated data (simulated stock returns follow a t-distribution with 15 degrees of freedom) in Table \ref{tab:backsimu}, both empirical and SVl-EVT models are rejected by $LR_{uc}$ and $LR_{cc}$ due to inappropriate exceedance counts. In contrast, the SVtl-EVT model performs best, not rejected by any tests, confirming exceedance numbers within the confidence interval.

\begin{table*}[h]
\centering
 \caption{Binomial, unconditional, conditional, and independence coverage tests based on different models using test data.  }
  \begin{tabular*}{2\columnwidth}{@{\extracolsep{\fill}}cccccc}
    \toprule
  \cmidrule(r){2-6}                  
& Binomial (95\%CI)  & Unconditional     & Independence     & Conditional  &  Exceedance\\
    \cmidrule(r){2-6}
& $[37,63]$  &  $LR_{uc}$     &$ LR_{ind}$     & $LR_{cc}$  &  \\
    \toprule
    SVt-EVT &*& 0 &  $8.958^{**}$    &$8.958^{**}$  & 50\\
    SVl-EVT    &* &1.081&$9.700^{**}$   &$10.781^{**}$& 43\\
    SVtl-EVT     &*&3.294& $3.406^{**} $ &$6.700^{**}$ &38\\
    GARCH-EVT&* &1.616&$ 12.457^{**}$  &$14.073^{**} $&59 \\
     Empirical    & &$6.161^{**}$&$24.854^{**}$   &$31.006^{**}$   & 68\\
      GARCH     &*&0.328&$ 9.657^{**}$  & $9.985^{**}$& 54 \\
    \bottomrule
  \end{tabular*}\label{tab:back}
\end{table*}

\begin{table*}[h]
\centering
 \caption{Binomial, unconditional, conditional, and independence coverage tests based on different models using simulated data. }
  \begin{tabular*}{2\columnwidth}{@{\extracolsep{\fill}}cccccc}
    \toprule
  \cmidrule(r){2-6}                  
 &Binomial (95\% CI) & Unconditional     & Independence     & Conditional  &  Exceedance\\
    \cmidrule(r){2-6}
& $[37,63]$  & $ LR_{uc} $    & $LR_{ind} $    & $LR_{cc}$  &  \\
    \toprule
   SVt-EVT & *& 2.747& 0.149   &2.896  &39 \\
    SVl-EVT    & &$15.994^{**}$& 1.284  &$17.278^{**}$ &25 \\
    SVtl-EVT    &* &1.810&0.064   &1.874 & 41\\
    GARCH-EVT&* &1.984& 1.538  &3.522 &60 \\
     Empirical    & &$4.345^{**}$&3.152   &$7.497^{**}$   &65 \\
      GARCH    &* &0 & 2.201&2.201 & 50 \\
    \bottomrule
  \end{tabular*}\label{tab:backsimu}
\end{table*}

\section{Conclusion}
In this paper, we introduce a methodology that integrates SV and GARCH models with Extreme Value Theory (EVT) to estimate and backtest Value-at-Risk (VaR). The extended SV models, SVt and SVtl, address time-varying volatility, fat tails, and leverage effects, with parameters estimated via the MCMC algorithm. The Peaks Over Threshold (POT) method from EVT, utilizing maximum likelihood estimation, effectively captures the tail distribution of residuals. Applied to the S\&P 500 and simulated returns, these models provide robust predictions of future returns and volatility. The SV-EVT models, particularly SVtl-EVT, demonstrate superior performance in VaR estimation compared to GARCH-based and empirical methods. Despite challenges in exceedance clustering, particularly during extreme events like the COVID-19 pandemic, the SVtl model shows the fewest exceedances. Further enhancements, including the adoption of SVJt and SVLJt (stochastic volatility model with leverage effect, fat-tail and ) models, are suggested for better handling of extreme scenarios.

\bibliographystyle{plainnat}
\bibliography{references.bib}
\end{document}

%% file: defs.tex
\usepackage{amsmath}
\usepackage{amsfonts}
\usepackage{mathrsfs}
\usepackage{bbm}

\def\calN{{\mathcal{N}}}

\def\X{{\mathbf{X}}}

\def\h{{\mathbf{h}}}

\def\x{{\mathbf{x}}}
\def\y{{\mathbf{y}}}

\usepackage{amsthm}
\usepackage{etoolbox}
\theoremstyle{plain}